\documentclass[5p,twocolumn]{elsarticle}

\usepackage{amsmath}
\usepackage{graphicx}   
\usepackage{breqn}
\usepackage{natbib}
\usepackage{amssymb}
\usepackage{multicol,caption}
\usepackage{multicol}

\begin{document}

\newcommand{\simtel}{\textit{sim\_telarray}}
\newcommand{\impact}{{ImPACT}}

\title{A Monte Carlo Template based analysis for Air-Cherenkov Arrays}

\author[rvt]{R.D. Parsons\corref{cor1}}
\author[uvl]{J.A. Hinton}
\cortext[cor1]{Corresponding author \\
Email Address: Daniel.Parsons@mpi-hd.mpg.de \\
Tel: +49 6221 516137}

\address[rvt]{Max-Planck-Institut f\"ur Kernphysik, P.O. Box 103980, D 69029, Heidelberg, Germany}
\address[uvl]{Department of Physics and Astronomy, The University of Leicester, University Road, Leicester, LE1 7RH, United Kingdom}

\date{\today}

\begin{abstract}
  
  We present a high-performance event reconstruction algorithm: an
  Image Pixel-wise fit for Atmospheric Cherenkov Telescopes
  (\impact{}).  The reconstruction algorithm is based around the
  likelihood fitting of camera pixel amplitudes to an expected image
  template. A maximum likelihood fit is performed to find
  the best-fit shower parameters.  A related reconstruction algorithm has
  already been shown to provide significant improvements over
  traditional reconstruction for both the CAT and
  H.E.S.S. experiments. We demonstrate a significant improvement to
  the template generation step of the procedure, by the use of a full
  Monte Carlo air shower simulation in combination with a ray-tracing
  optics simulation to more accurately model the expected camera
  images.  This reconstruction step is combined with an MVA-based
  background rejection.

Examples are shown of the performance of the \impact{} analysis
on both simulated and measured (from a strong VHE source) gamma-ray
data from the H.E.S.S. array, demonstrating an improvement in sensitivity of more than a
factor two in observation time over traditional image moments-fitting methods,
with comparable performance to previous likelihood fitting analyses. \impact{} is a
particularly promising approach for future large arrays such as the
Cherenkov Telescope Array (CTA) due to its improved high-energy
performance and suitability for arrays of mixed telescope types.

\end{abstract}

\begin{keyword}
Gamma-ray astronomy; IACT; Analysis technique; Cherenkov technique
\end{keyword}

\maketitle

\section{Introduction}

Ground-based gamma-ray astronomy exploits the air shower produced by the
interaction of a primary gamma ray in the Earth's atmosphere. For gamma-rays of
approximately $10^{11} - 10^{14}$ eV (VHE) few air-shower particles reach ground level. 
However, sufficiently high-energy secondary particles will emit Cherenkov radiation,
resulting in illumination of a $\sim$10$^{5}$ m$^{2}$ patch of ground for a few nanoseconds.
The Imaging Atmospheric Cherenkov Technique (IACT) is based on the 
observation of this emission by a number of large reflecting
telescopes, placed within the light pool with a typical spacing of $\sim$100~m.
The Cherenkov light is focussed onto ultra-fast, but typically rather coarsely pixelated, cameras,
resulting in roughly elliptical images of the
shower emission in multiple telescopes. H.E.S.S. \citep{Hinton2004}
is an array of four such 100\,m$^{2}$ reflecting telescopes, at 1800\,m altitude
in the Khomas Highlands of Namibia, each with a 960 pixel camera 
viewing a 5$^\circ$ diameter patch of sky.

The process of reconstruction/estimation of the properties of the primary photon (direction and energy), 
and the rejection of events likely to belong to the  background of hadronic showers,  makes use of image information from each telescope.
Traditionally this event reconstruction is performed using the \textit{Hillas} parameters
of the camera images~\cite{Hillas1985}, derived after an image
cleaning step (described in \cite{Aharonian2006}). The Hillas parameters are the moments
of the camera image which, given the approximately elliptical nature
of typical camera images, already capture much of the available image information. 
In the most commonly used stereoscopic reconstruction method the major axes of images are calculated 
in a common camera reference frame and the intersection points of all axes found. 
A weighted average (based on image amplitude and the angle between the axes) is then taken of all 
crossing points to provide an estimate of the arrival direction of the
primary gamma-ray. A similar procedure involving the intersections of the directions between the image centroid and the optical axis is then performed in
a common plane perpendicular to the pointing direction, to determine the shower impact point on the ground.

Although relatively good angular resolution ($\sim$0.1$^\circ$ with H.E.S.S.) 
can be reached with such a reconstruction procedure~\citep{Aharonian2006}, significant 
additional information can be extracted from the recorded images in a typical event, resulting in improved performance.
Additionally a simultaneous fit procedure between all telescopes can help to ensure
a consistent result is found between all telescopes, rather than the independent analysis
used in Hillas reconstruction.
An image template fitting procedure was pioneered for the CAT 
telescope \citep{Barrau1998} \cite{LeBohec1998}
and re-implemented and improved upon for H.E.S.S. \cite{deNaurois2009}.
These methods begin with the creation of  
a semi-analytical model of air-shower development
and IACT response, and use this model to generate the expected shower image for a given set of shower 
parameters (primary energy and direction, and also impact distance of the shower from the telescope). 
The template library can then be compared to the images recorded in a given event, and, 
by means of a multi-dimensional fit procedure, the best-fit shower parameters determined.
An alternative method of shower fitting has also been developed for use with H.E.S.S. data, fitting the pixel
intensities of the camera image with the expectation from a simpler analytical 3 dimensional
gaussian air shower model (3D model) \citep{Lemoine2006,Becherini2011}.

One of the major problems with the \textit{model} and 3D model analyses is the 
difficulty of describing the air shower behaviour at high energies ($>$10\,TeV). Above 10\,TeV
a large number of particles reach ground level, resulting in large fluctuations which are difficult
to reproduce with the aforementioned approaches. This difficultly in reproducing energetic air showers
results in poor event reconstruction above 10\,TeV, typically leading to a rapid drop in effective collection area.
With these approaches it is also difficult to account for instrumental effects, such as
the optical point spread function of the telescopes or the limited readout window of the camera, so
approximations of these effects must be made. Inevitably, making these kind of assumptions
will limit the quality of the model fit, reducing the accuracy of the analysis.

One way to solve these problems with high energy events is to instead
produce templates by the use of detailed Monte Carlo (MC) based air shower simulations. Such
an approach requires no analytical approximations to be made within the air shower simulation,
and therefore takes better into account the large fluctuations arising from particles reaching ground level. 
Additionally, a more accurate optics and electronics simulation can be performed on the simulated Cherenkov photons, resulting
a more realistic representation of the telescope behaviour.  The MC approach is a robust and
general image template generation method, not requiring any adaptation to the model for different
telescope types, instead existing Monte Carlo telescope configurations can be adapted to produce 
templates.

Below we present an attempt to improve the accuracy of the template generation, by the use of
a more accurate Monte Carlo simulation based air shower model, combined with a ray-tracing telescope simulation
and demonstrate the resultant improvement in the performance of the analysis.

 \begin{figure*}[t]
\begin{center}
\includegraphics[width=0.99\textwidth]{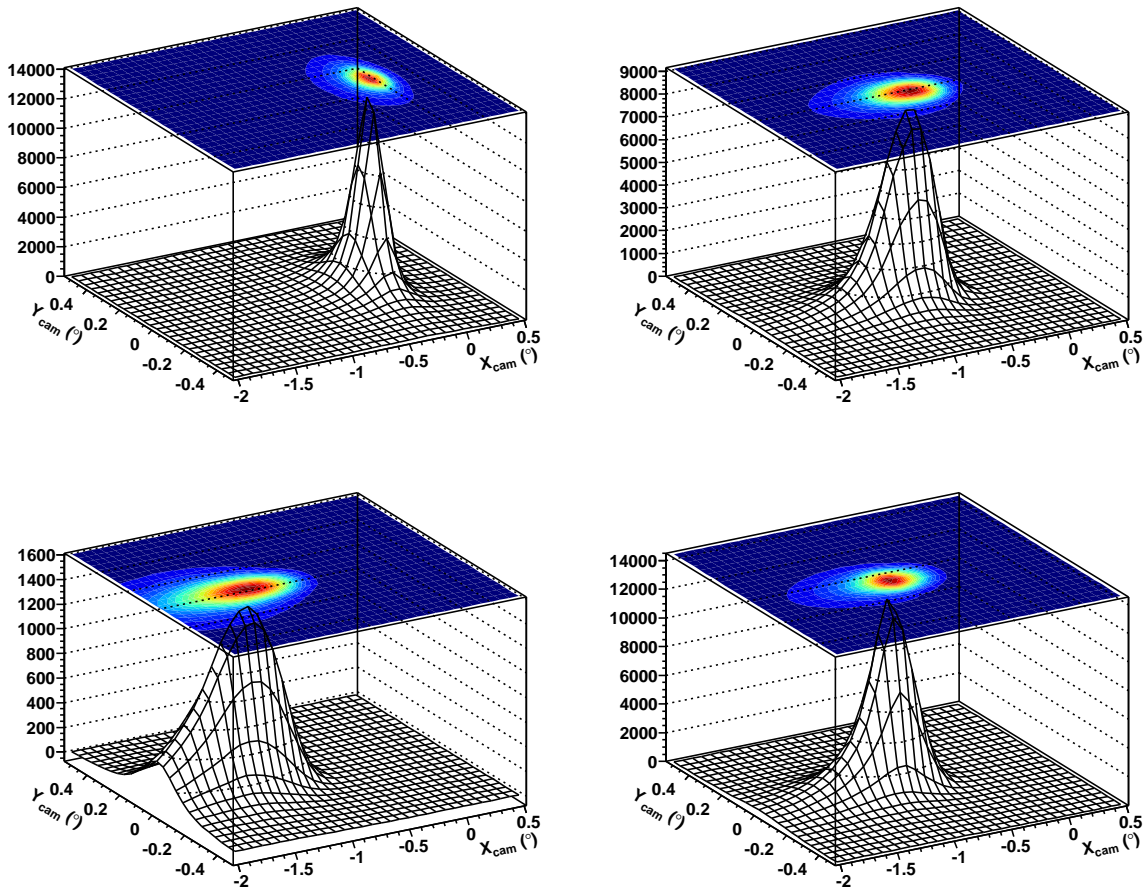}

\caption{Image template histograms for a 1\,TeV primary gamma-ray
at a core distance of 20\,m (top left), 100\,m (top right) and 200\,m (bottom left)
at the expected shower maximum (300 g\,cm$^{-2}$). The bottom right plot
shows the shower template at a core distance of 100\,m and a X$_{\rm max}$
of 400 g\,cm$^{-2}$. The $z$-axis is shown in units of photoelectrons per square 
degree, and the $x-$ and $y-$axes are in degrees.}
\label{fig-templates}
\end{center}
\end{figure*}

\section{Template Generation}

In contrast to the semi-analytical model of shower development used in \cite{LeBohec1998} \& \cite{deNaurois2009}
to generate image templates, we use instead the CORSIKA/\simtel{} chain \citep{Heck1998,Bernlohr2008}
to perform a Monte Carlo air shower simulation, followed by
ray-tracing of the telescope optics
and simulation of the instrument electronics.
The use of this very detailed and computing intensive simulation chain has several major benefits over the,
 faster, semi-analytical approach.

Firstly, the CORSIKA air shower simulation is a well proven code, which has been shown to be consistent with the 
results of multiple experiments, in combination with the \simtel{} telescope simulation it has been shown to
accurately reproduce results from H.E.S.S..
The use of a complete air shower simulation also allows the effects of
the geomagnetic field on the air shower to be incorporated,
which proves difficult in  analytical models. Finally the simulation of the electronics chain using \simtel{}
allows instrumental effects, such as trigger biases and readout window
length (very important in the case of H.E.S.S.) to be naturally accounted for.

There are, however, some drawbacks to the MC approach, firstly, in order to fill the same parameter
space of templates, a significantly larger computing time is required, due to the CPU intensive
nature of MC simulations. Also, with such a random approach the edges of the parameters space are 
naturally sparsely populated, making the fitting of unusual events
more difficult. This issue of phase space coverage was most apparent close to the array
trigger threshold, where an extremely large number of event simulations were required in order to
reproduce the relatively few up-fluctuations that trigger the array.

Full MC simulation of a number 
of air showers (300,000 -- 600 depending on the primary gamma-ray energy) 
were performed for a given event parameter set and 
traced through the simulation of the optics. Any photons reaching the focal
plane in an event that triggers the camera were then given a weight 
in order to account for a number of instrumental
effects and efficiencies (e.g. mirror reflectivity, quantum efficiency, integration time 
window), such that they represent a number of detected photoelectrons (p.e.).
The average image for a given set of shower parameters was then saved 
in the form of a finely binned histogram, containing
the expected image in a `perfect' camera. 
These histograms were then oversampled with the camera pixel 
size, producing lookup tables for the expected image amplitude at all 
pixel positions within the camera (see Figure~\ref{fig-templates}).

This procedure was repeated at points in a four-dimensional parameter space (described below)
to produce a comprehensive library of image templates. 

\begin{itemize}
  \setlength{\itemsep}{-2pt}
\item 8 zenith angles (0-65$^\circ$)
\item 2 azimuth angles 
\item 17 energies (0.08-100\,TeV)/$\cos(\rm{zen})$
\item 25 impact distances (0-1000\,m)
\end{itemize}

Additionally events were binned in a number of bins of X$_{\max}$, of
width 25 g\,cm$^{-2}$ spread
around the expected X$_{\rm max}$ for a shower of that energy. In total such a 
scheme should produce over 100,000 image templates, however in practice the 
number was somewhat smaller (less than half this total), as templates with insufficient 
Cherenkov photon statistics were discarded. 
A multidimensional linear interpolation algorithm was then used
to interpolate between these 
templates, allowing an expected image template
to be produced for any shower parameters within the above ranges.

An additional parameter ignored here is the position of the gamma-ray source
within the camera field of view, which could be important due to the broadening of the
telescopes optical point spread function (PSF) with distance from the camera centre.
However for H.E.S.S., with its Davies-Cotton optics \citep{Davies1957}, the PSF does 
not degrade significantly across the field of view, so events simulated on axis are 
sufficient in most cases.

\section{Likelihood Fitting}

 \begin{figure}[t]
\begin{center}

\includegraphics[width=0.49\textwidth]{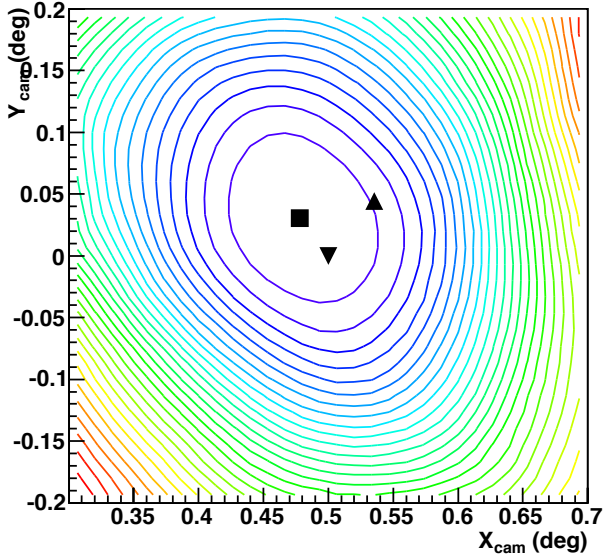}

\caption{2D projection of a likelihood surface for an example
  event in the plane of the camera. 
The triangle shows the reconstructed shower direction derived using a traditional 
Hillas-based reconstruction, the square shows the direction as reconstructed by 
\impact{} and the inverted triangle shows the true direction of the simulated shower. }
\label{fig-likesurf}
\end{center}
\end{figure}

Once the full set of templates for the range of possible shower parameters has been
created, they must then be compared with the observed image and a fit performed
to find the shower parameters that best fit the observed event.

The comparison between the expected and measured images is made using the 
likelihood function developed in \cite{deNaurois2009}. This likelihood of a signal $s$ given an
expectation of $\mu$ consists of a convolution of 
the Poisson distribution of each individual photoelectron $n$, with the resolution of the 
photosensor. Where the resolution of the photosensor is represented by a Gaussian
of width $\sqrt{\sigma_p^2 + n\sigma_\gamma^2}$, where $\sigma_p$ is the width of
the pedestal (charge distribution from night sky background light and electronic
noise) and
$\sigma_\gamma$ is the width of the single photoelectron distribution ($\approx$0.5 in H.E.S.S.).

\begin{multline}
P(s|\mu,\sigma_p,\sigma_\gamma) = \sum_{n} \frac{\mu^n e^{-\mu}}{n! \sqrt{2 \pi (\sigma_p^2 + n\sigma_\gamma^2)} }  \\ \cdot \exp \left( - \frac{(s-n)^2}{2 (\sigma_p^2 + n\sigma_\gamma^2)} \right)
\label{eqn-like}
\end{multline}

The log-likelihood for that pixel is then defined such that it is distributed similarly
to a $\chi^2$ distribution.

\begin{equation}
\ln L = -2 \ln P(s|\mu,\sigma_p,\sigma_\gamma)
\label{eqn-logl}
\end{equation}

In cases of large signal expectation ($\mu>5$) the Poissonian signal fluctuations can be replaced
by a Gaussian, eliminating the need to perform the sum over all photoelectrons.

\begin{multline}
P(s|\mu \gg 0,\sigma_p,\sigma_\gamma) \approx \frac{1}{\sqrt{2 \pi (\sigma_p^2 + \mu (1 + \sigma_\gamma^2))} } \\ \cdot \exp \left( - \frac{(s-\mu)^2}{2 (\sigma_p^2 + \mu(1 + \sigma_\gamma^2))} \right)
\label{eqn-likeGaus}
\end{multline}

More detailed derivations of equations \ref{eqn-like} \& \ref{eqn-likeGaus} can be found
in \cite{deNaurois2009}.

It should be noted that when calculating the likelihood that two different gain channels
are used within the H.E.S.S. camera. The high gain channel covering the range
of $\sim$0-150 p.e. and the low gain channel used above.
These two channels
have different levels of electronic noise and hence different pedestal widths (around 0.2 p.e. for 
low gain and 1 p.e. for high gain \cite{Aharonian2004}) these pedestal values are broadened further
during observations by the presence of night sky background light. However, in practice the
contribution of the pedestal to the low gain channel (used for large signals) is small, only 
having an effect in pixels where the high gain channel is
unavailable. 

Once this per pixel likelihood function has been defined it can be simply
summed over all significant pixels, selected based on the two level tail cut described in \cite{Aharonian2006},
with two additional rows of pixels added around the image edge,
and summed over all telescopes passing selection cuts to find an event likelihood for
a given set of shower parameters. 
This event likelihood must then be minimised  
in a 6-dimensional fit over direction, impact point, $X_{\rm max}$ and
primary energy. In order to simplify the reconstruction of $X_{\rm max}$ it
is first reconstructed using a geometrical approach, assuming depth of maximum
corresponds to the brightest point in the image (calculated by taking the average position
of the brightest 3 camera pixels).
The minimisation can then be performed over a modification factor to 
the estimated $X_{\rm max}$, greatly reducing the time taken for the fit procedure.

Fitting is performed using the widely-used MINUIT \citep{James1975} package,
providing 
a fast and reliable minimisation. 
The algorithm finds a function minimum in the majority of
cases, typically taking around 500 function calls to reach the minimum, with a computation time of 
$\sim$0.2-0.5 seconds. 

\begin{figure*}[]
\begin{center}
\includegraphics[width=0.49\textwidth]{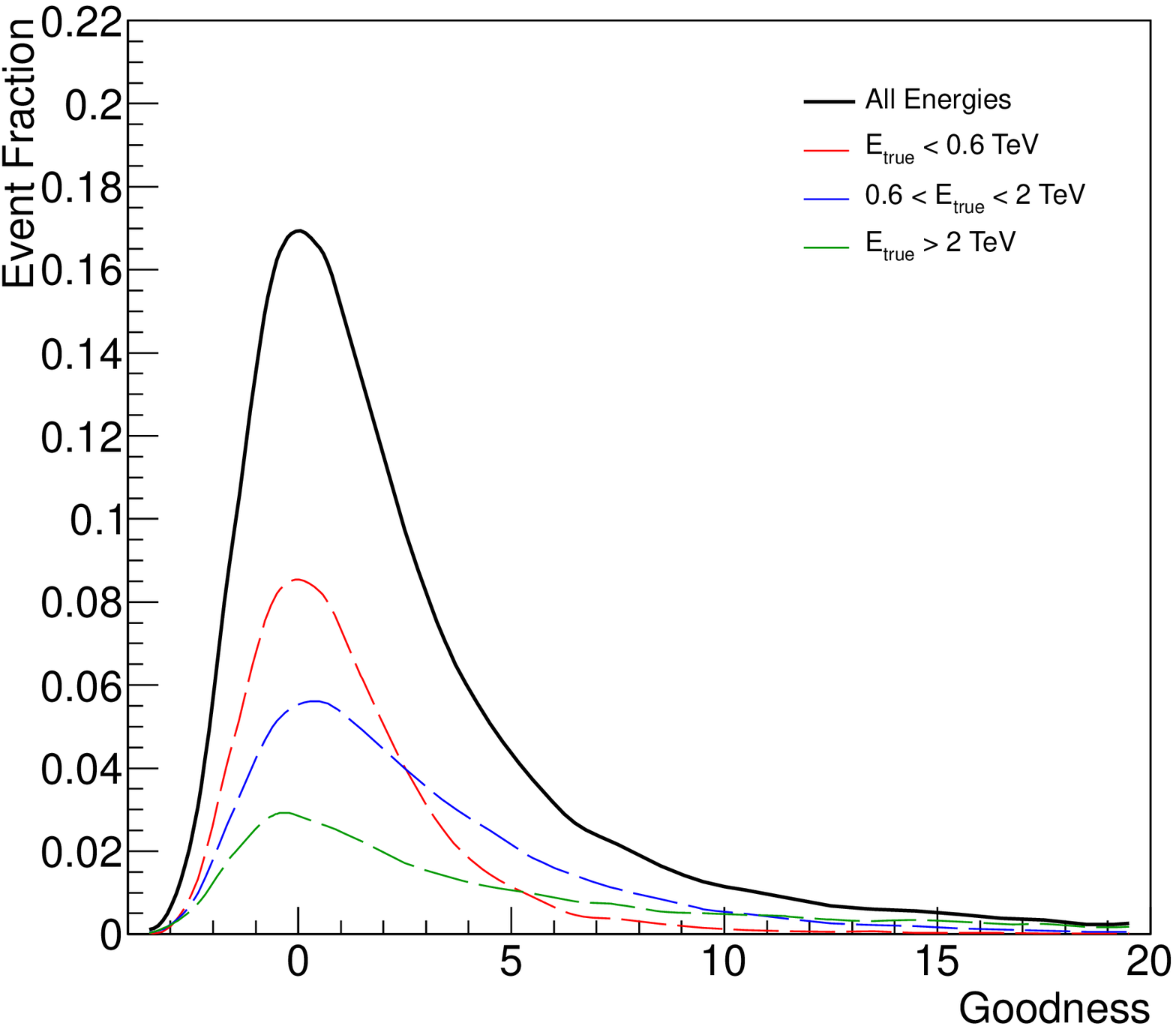}
\includegraphics[width=0.49\textwidth]{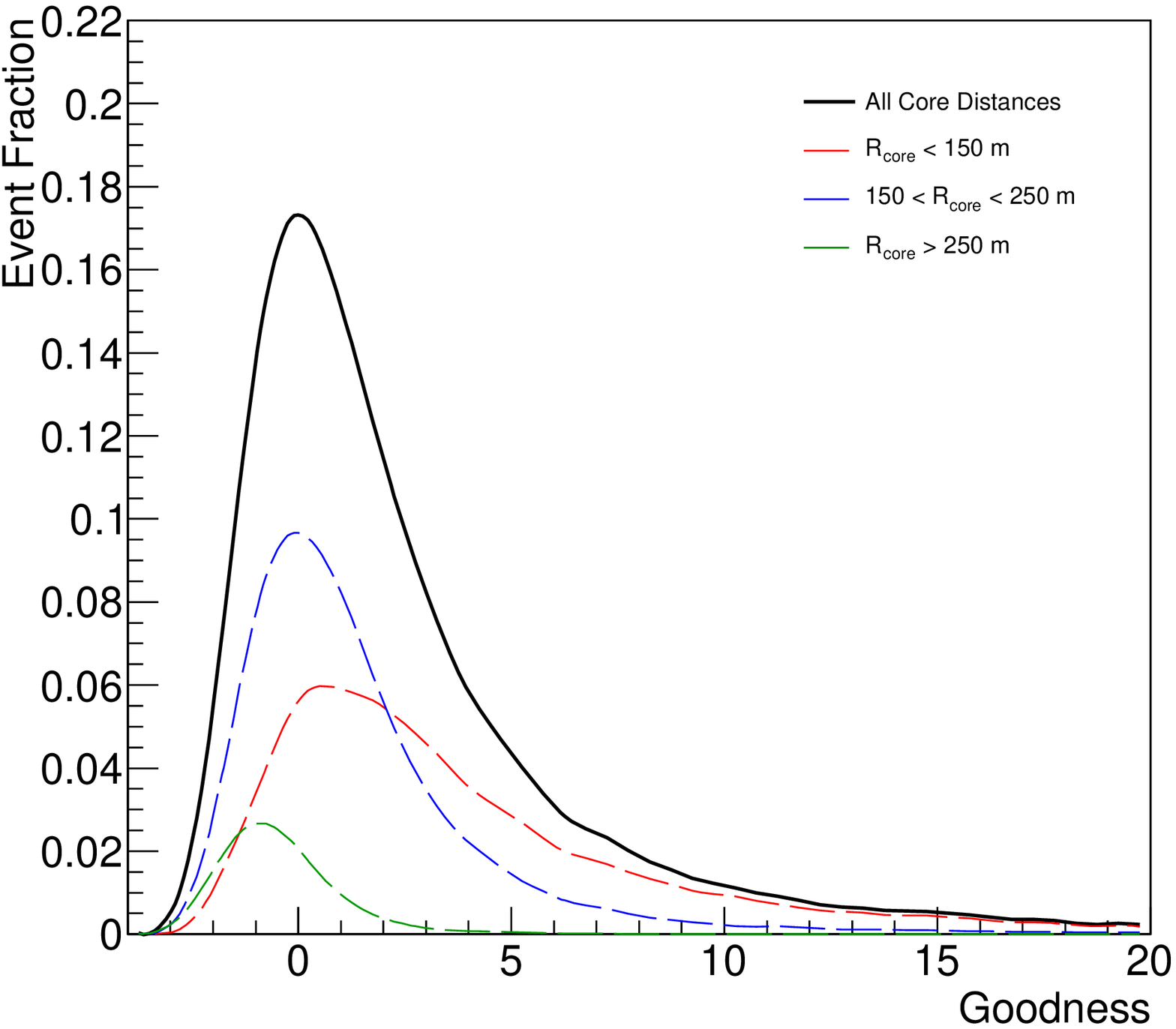}

\caption{Distribution of goodness-of-fit for pixels with an expected signal
of over 0.01 p.e. for simulated gamma-rays
split into bins of true (simulated) energy (left) and bins of simulated 
distance of the shower core from the telescope (right). Plots are made
 at 20$^\circ$ zenith angle. }
\label{Gdist}
\end{center}
\end{figure*}

With such a high dimensionality, using an appropriate seed position for the fit
is crucial to avoid getting trapped in a local minimum. Multiple Hillas-style
reconstructions using different image cleaning values are used to generate
possible fit seed positions, in addition to a the axis reconstructed 
using the \textit{disp} method \cite{Hofmann1999,Lu2013}. 
Several steps are then made using
a simple steepest descent algorithm from each seed position
to allow the fit to approach the closest minimum 
and the position with the highest likelihood used to seed the MINUIT fit.

Figure \ref{fig-likesurf} shows a 2 dimensional projection of the likelihood surface 
in the camera plane. This example event demonstrates the smoothness and well 
defined minimum of the likelihood surface. In this case the MIGRAD algorithm is able 
to find the minimum point in this function at a more accurate position than the
Hillas-based reconstruction.

\subsection{Goodness of Fit}

In order to create a useful goodness of fit statistic it is necessary to know the
average behaviour of the likelihood function, so that a comparison to the 
resultant log-likelihood of the fit can be made. The average log-likelihood for a 
given $\mu$, $ \sigma_p$ and $\sigma_\gamma$ can be calculated as below:

\begin{equation}
\langle \ln L \rangle |_\mu = \int ds \ln L(s|\mu, \sigma_p,\sigma_\gamma) \times P(s|\mu,\sigma_p,\sigma_\gamma).
\label{eqn-likeAv}
\end{equation}

Again in the case of large signals this formula can be greatly simplified to its Gaussian
limit.

\begin{equation}
\langle \ln L \rangle |_\mu = 1 + \ln(2 \pi) + \ln(\sigma_p^2 + \mu (1+\sigma_\gamma^2) )
\label{eqn-likeAvGaus}
\end{equation}

More detailed derivations of equations \ref{eqn-likeAv} \& \ref{eqn-likeAvGaus} can be found
in \cite{deNaurois2009}.
As equation  \ref{eqn-likeAv} contains an integral with no analytical solution it would be
expensive to compute this value on the fly, instead it is preferable to compute this value before the analysis,
storing the values in lookup tables as the difference from the Gaussian expectation. This average
behaviour can then be compared with the fit result to produce a goodness-of-fit statistic ($G$).

\begin{equation}
G = \frac{ \sum\limits_{i} [ \ln L(s_i | \mu_i)  - \langle \ln L \rangle |_\mu  ] }{\sqrt{ 2 \times  \rm{NdF}} }
\end{equation}

This value is constructed such that if all pixels behave like independent random 
variables the resultant distribution should be Gaussian in form and centred at 0.
However, figure \ref{Gdist} shows that this is not the case, with a long tail of high G
values being seen at high energies and small impact distance. This tail originates from high-energy showers at 
small impact distance, where visible sub-structure within the shower produces correlated 
fluctuations in multiple pixels.
It should also be noted that the peak of the G distribution is affected
by the level of night sky background (NSB).

\begin{table}[b]
\begin{center}
\begin{tabular}{|c|c|c|c|}

\hline
Config & Amp (p.e.)& $\theta^2$ (deg$^2$)& $\zeta$ \\
\hline
{Hillas Std } & 60 & 0.0125 & 0.84 \\
{Hillas Hard} & 160 & 0.01 & 0.89\\
{ImPACT Std} & 60 & 0.005 & 0.83 \\
\hline

\end{tabular}
\end{center}
\caption{Image selection and background rejection cuts for the three H.E.S.S.
cut configurations compared in this paper.}
\label{tab-cuts}
\end{table}%

\section{Background Rejection}

The discrimination of the gamma-ray induced air showers from the
much more numerous cosmic-ray induced showers is an important factor in
maximising the sensitivity of an
 analysis. One option for background rejection
is to use the aforementioned goodness-of-fit value. However, due to the strong 
dependence of this value of the NSB level (discussed in detail in \citep{Becherini2011} \& \citep{deNaurois2009}) 
and the good knowledge of the single p.e. and pedestal width required, this goodness of fit value
may not be stable between different observations conditions.
Although this value will be a powerful background discrimination parameter, in this work it has not 
been used as the primary means of background rejection. However, with further work on calibration 
between observation conditions both 
this goodness of fit value and the estimated error on the direction reconstruction may be 
very useful in the rejection of cosmic-ray events.

\begin{figure*}[]
\begin{center}
\includegraphics[width=0.49\textwidth]{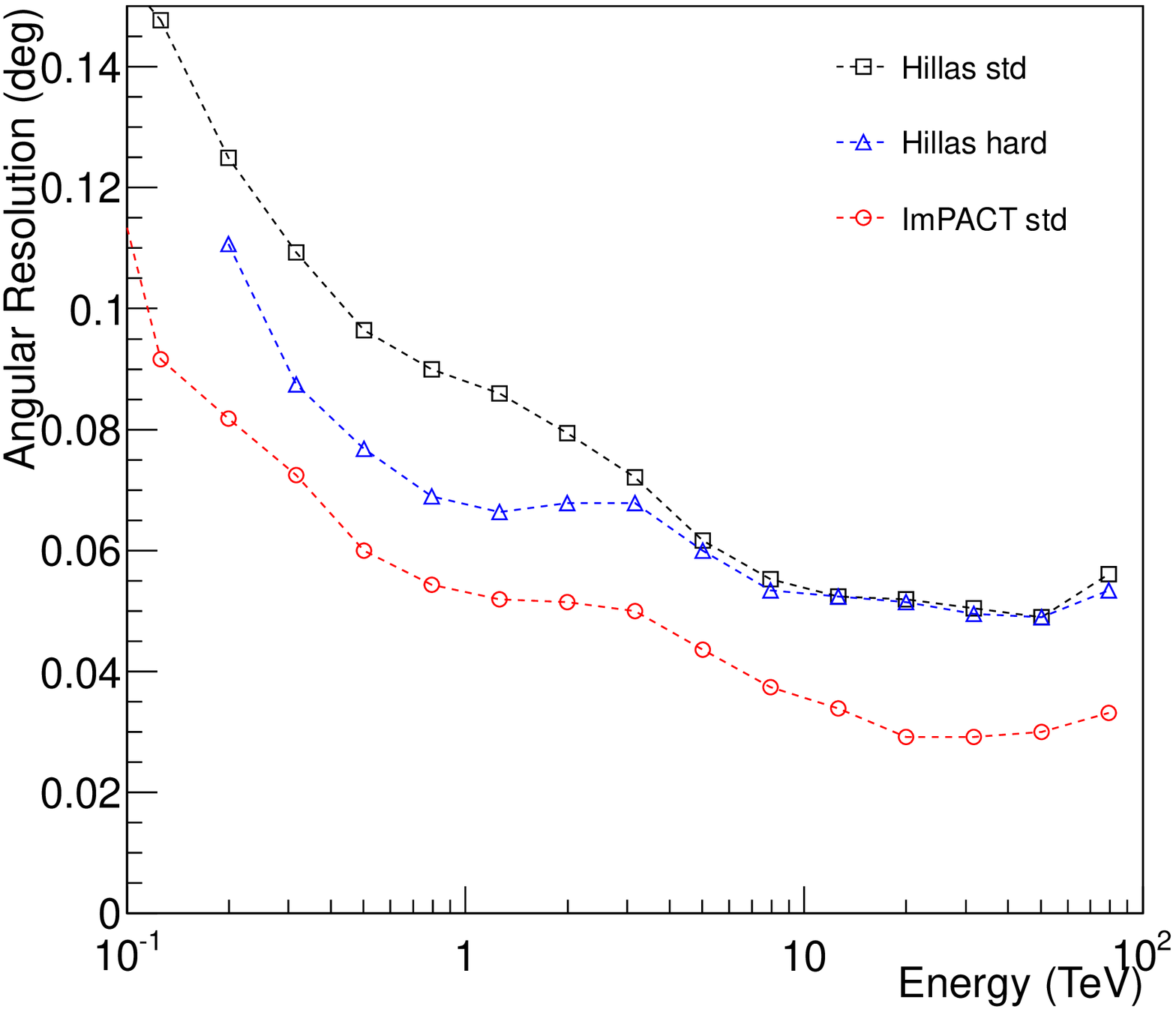}
\includegraphics[width=0.49\textwidth]{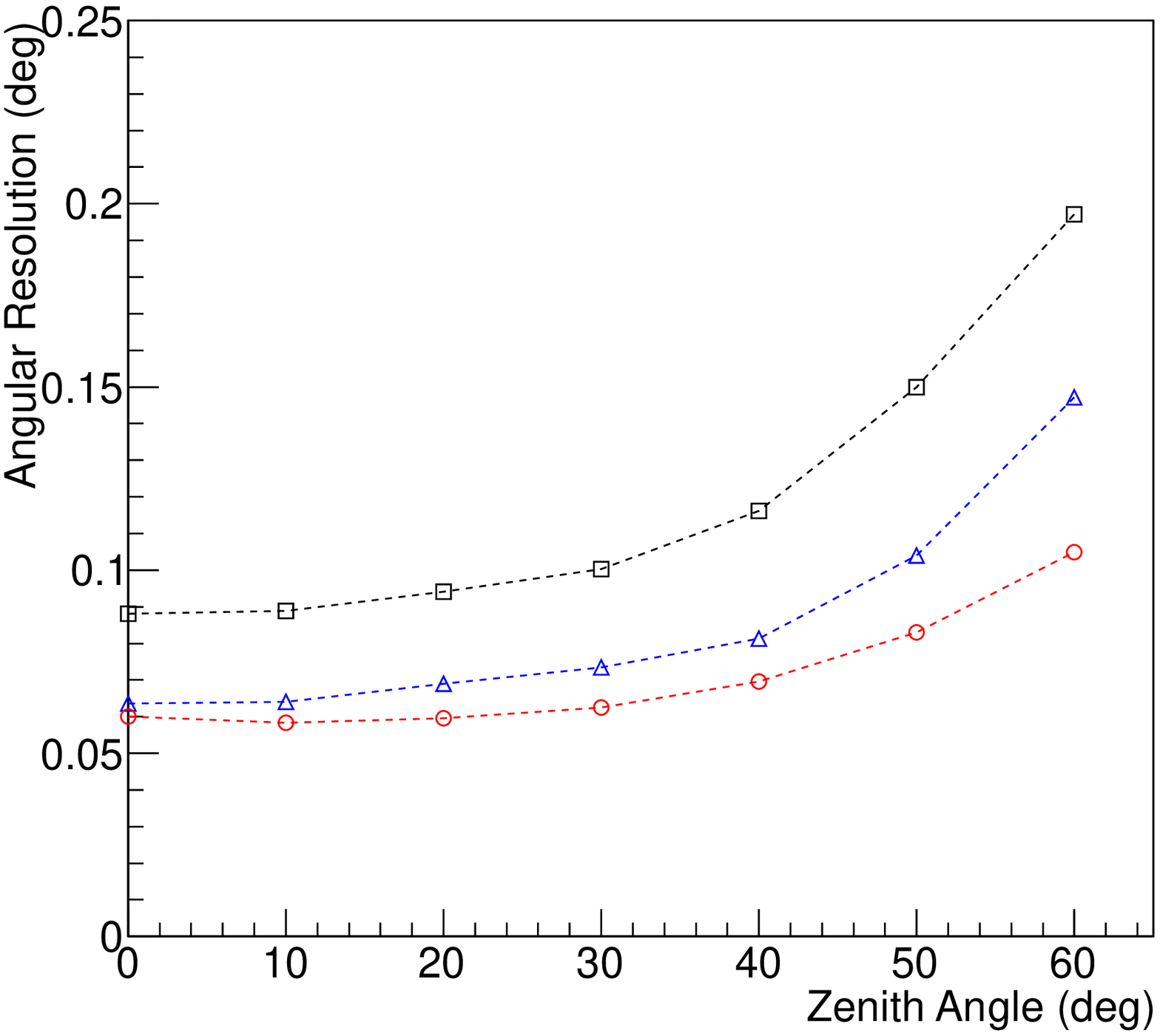}
\caption{Angular resolution (68\% containment radius) of the \impact{}
  method versus simulated energy at $20^{\circ}$ zenith angle (left)
and versus zenith angle (right), averaging over a spectrum of the form
$dN/dE\propto E^{-2}$. Results are shown in comparison to
the \textit{Hillas std} and \textit{hard} results.}
\label{fig-angZen}
\end{center}
\end{figure*}

Background rejection is instead performed using the boosted decision tree (BDT) based
method already implemented within the H.E.S.S. framework (\citep{Ohm2009} and references therein)
using the root based TMVA package \citep{Hoecker2007}. This method combines a series
of discriminant parameters determined using a Hillas-based event reconstruction. These parameters
classify events based on comparison of the image widths and lengths to the expected mean values 
for both gamma-rays and hadron-induced showers, the spread in energy estimates from each telescope
and the reconstructed height of maximum of the air shower. The distribution of the discriminant parameters 
obtained on gamma-ray simulations and "off"-runs (observation runs with no strong gamma-ray emission in
the field of view) are then used to train a series of BDTs, created in a series 
of energy an zenith angle bins. Use of this multivariate analysis technique allows correlations of 
discriminant parameters to be effectively accounted for, producing a much more discriminating
parameter ($\zeta$) when compared to the 6 parameters independently.

Minimum amplitude (based on cleaned images), angular distance and background rejection cuts
were optimised for a series of point sources with different power-law spectra.
\textit{Standard cuts} are optimised for a source of 10\% Crab strength with a spectrum 
of E$^{-2.6}$ and \textit{hard cuts} are optimised for a source of 1\% Crab strength with a spectrum 
of E$^{-2}$. 
Additionally a 2$^\circ$ cut on the position of the image centroid from the camera
centre was applied, in order to remove highly truncated shower images.
Table \ref{tab-cuts} shows a comparison of the cut values for the currently used 
Hillas TMVA cut configurations in comparison with the ImPACT standard cuts.

Currently the \impact{} reconstruction uses boosted decision trees based only on the parameters 
extracted from the Hillas-style reconstruction of events. However, in the future it
may be possible to extract more information from the image template fit which can increase 
the background rejection power.

\section{Performance}

In this section the performance of the \impact{} reconstruction is compared
with the performance of the standard Hillas-style reconstruction with MVA-based background rejection \citep{Ohm2009}.
For brevitys
sake the results of only the \impact{} method standard cuts (hereafter referred to
as \textit{\impact{} std}) will be compared with both the standard and hard cuts
for the Hillas-style reconstruction (hereafter referred to
as \textit{Hillas std} and \textit{Hillas hard}).

The performance is assessed using Monte Carlo simulations of the 4 telescope H.E.S.S.
IACT array, with mirrors at 70\% of their nominal optical efficiency (roughly the
current state of the array).

\subsection{Angular Resolution}

\begin{figure*}[]
\begin{center}
\includegraphics[width=0.49\textwidth]{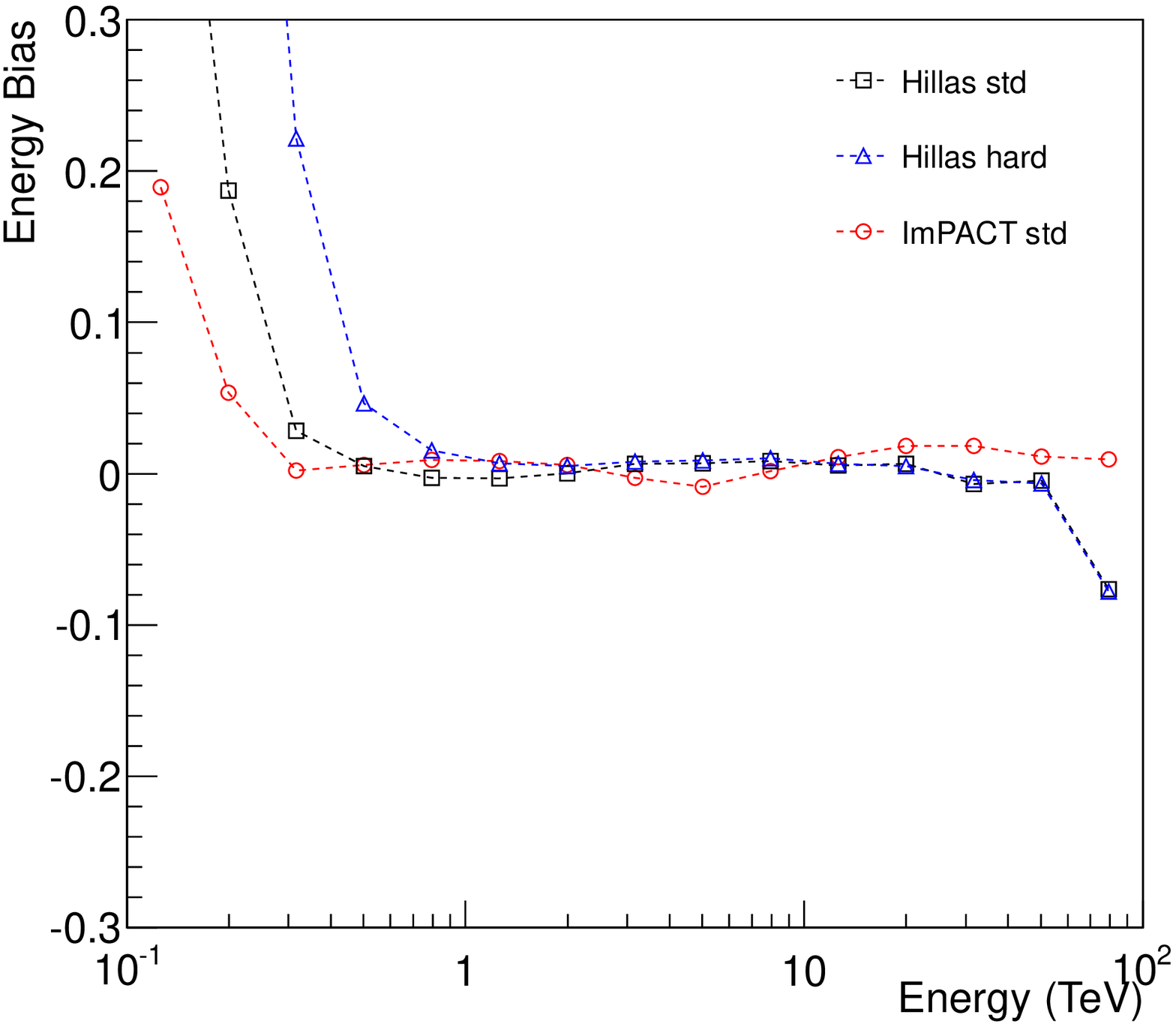}
\includegraphics[width=0.49\textwidth]{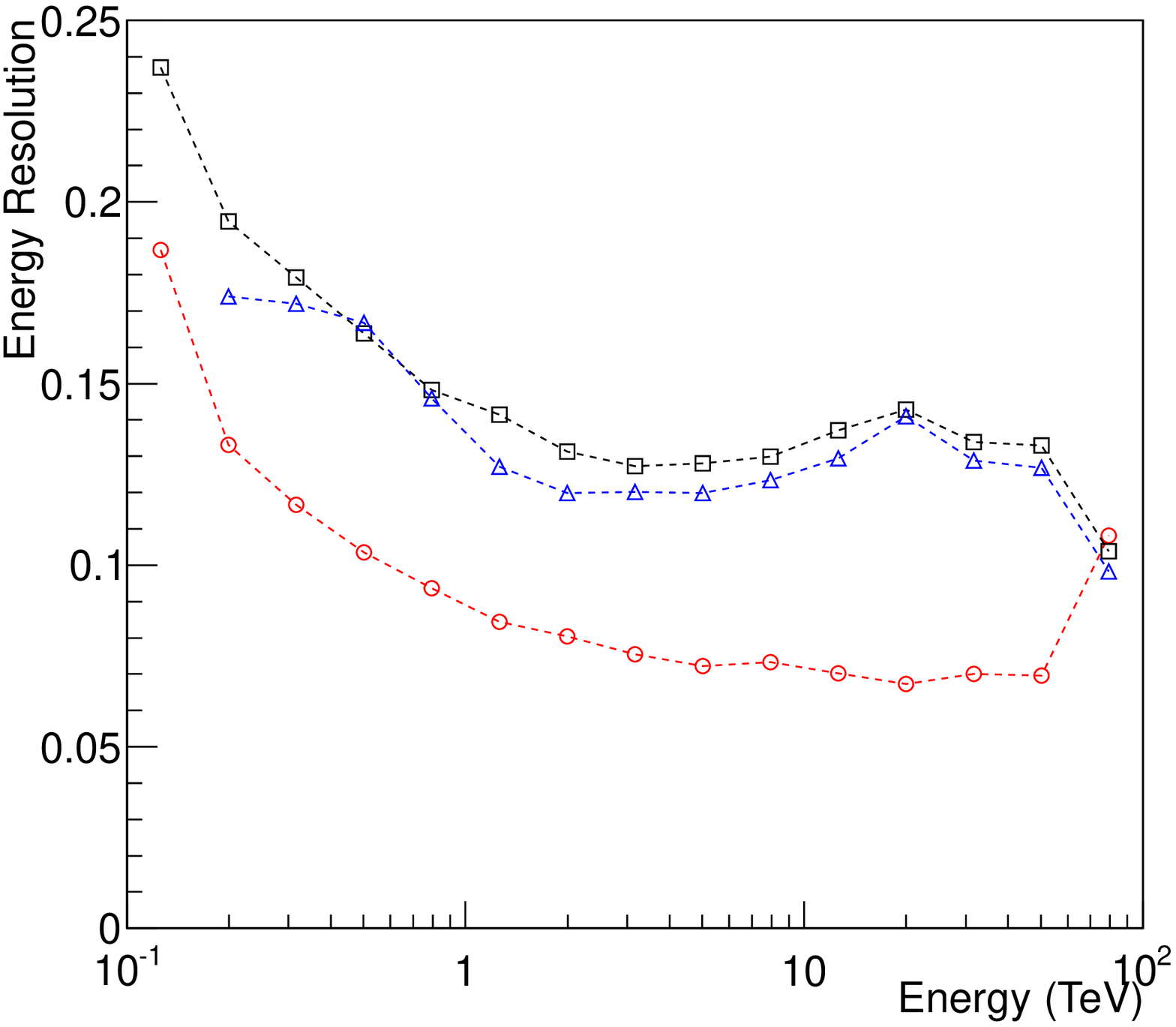}
\caption{Fractional energy bias (left) and energy resolution (right) for the \impact{} and
\textit{Hillas} methods, shown as a function of simulated photon energy.}
\label{fig-Eres}
\end{center}
\end{figure*}

The angular resolution, in this case defined as the 68\% containment radius of
the reconstructed event positions from a point-like source, is shown as a function
of energy and zenith angle in figure \ref{fig-angZen}. Figure 
 \ref{fig-angZen} (left) shows a clear improvement in angle resolution over the \textit{Hillas std}
 across the whole energy range.
This improvement ranges from around 50\% at 500\,GeV where the added
information has the largest effect, reducing to around 15\% at 100\,TeV where a
great deal of information is already provided by the Hillas parameterisation so the extra
information in the fit has less impact.

An improvement is also seen in figure \ref{fig-angZen} (right) at all zenith angles, this
demonstrates the weak dependence of the angular resolution on zenith angle for the 
\impact{} method, in contrast to the steep decline seen for the \textit{Hillas std}
method at high zenith angles. The rapid reduction of performance with zenith 
angle for the \textit{Hillas std} method, is due to the apparent foreshortening in the 
array in the frame of the shower at large zenith. This foreshortening results in more 
parallel camera images in high zenith events, making event reconstruction by image axis
intersection more difficult. The \impact{} method, however, considers more information
than just the image axis and copes better with this problem.

The aforementioned improvement in angular resolution will help in the study of
structure within large extended sources, allowing their study at smaller
angular scales. Smaller angular resolution also improves the sensitivity 
when observing point-like sources, as a smaller integration region around the source
position can be used, reducing the cosmic-ray background contamination.

\subsection{Energy Resolution}

\begin{figure*}[t!]
\begin{center}
\includegraphics[width=0.49\textwidth]{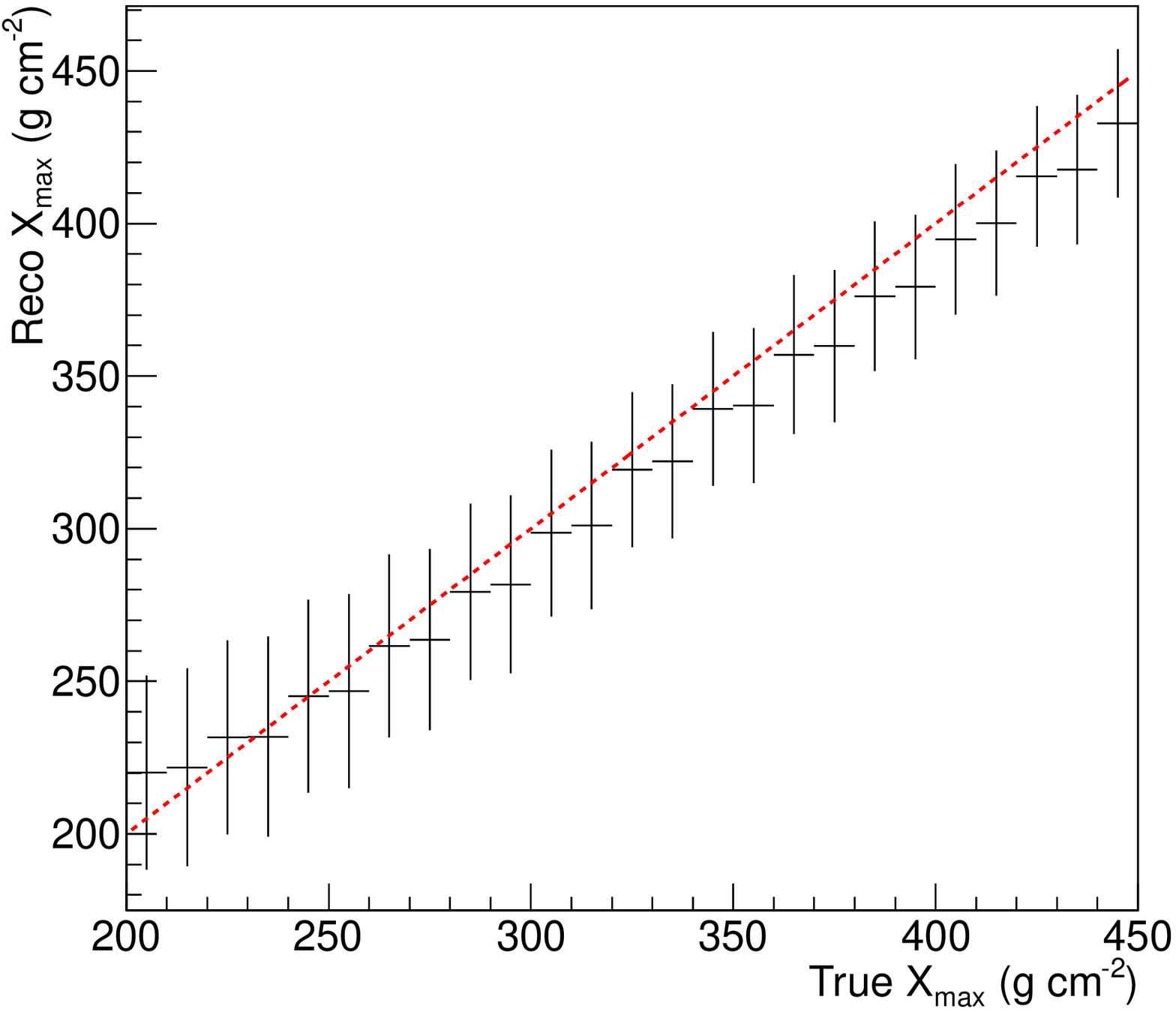}
\includegraphics[width=0.49\textwidth]{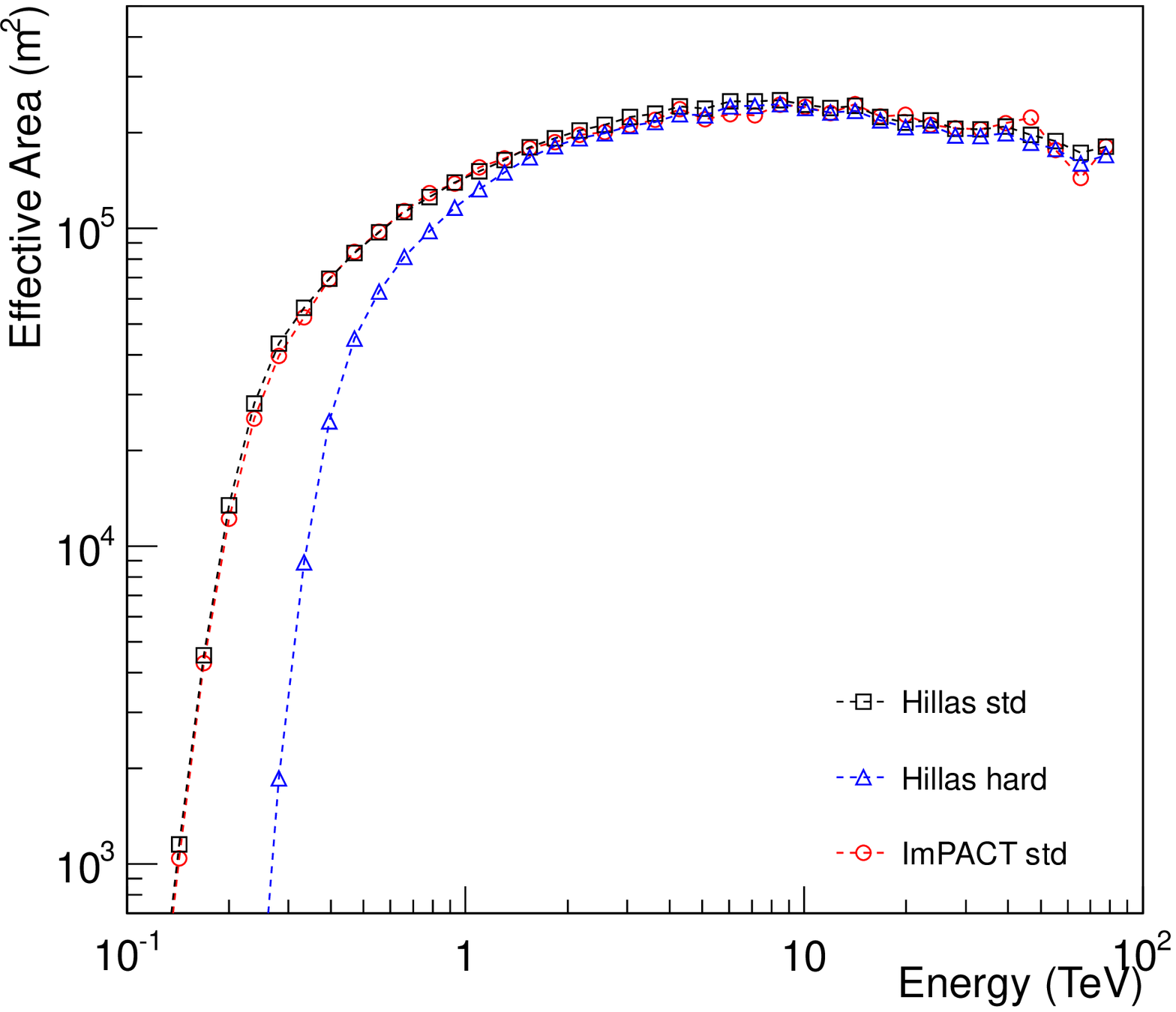}

\caption{Left: Average reconstructed X$_{\rm max}$ shown as a function of
simulated X$_{\rm max}$ for the \impact{} method, error bars represent the 
RMS of the reconstructed X$_{\rm max}$ distribution. The dashed red line shows
the line reconstructed X$_{\rm max}$ = simulated X$_{\rm max}$. Right: 
Effective collection area of the array as a function of energy for the
three sets of cuts/reconstruction methods considered.}
\label{fig-Xmax}
\end{center}
\end{figure*}

In addition to improving in the angular resolution of the array, the 
\impact{} method is also able to more accurately reconstruct the energy
of the primary air shower. The accuracy of reconstruction is often described by 
looking at the fractional deviation of the reconstructed energy from the simulated 
energy, figure \ref{fig-Eres} shows the mean (energy bias) and RMS (energy resolution) 
of this distribution as a function of energy.

The energy bias of all 3 reconstruction method is quite similar, all curves show
a large bias at low energies, where event selection cuts result in only extreme examples 
of low energy events (with large upward fluctuations in the number of Cherenkov photons
produced) being selected, which are subsequently poorly reconstructed.
For the \impact{} analysis, this low energy bias is slightly less extreme than the 
\textit{Hillas} reconstruction, resulting in a larger region of ``safe" energy reconstruction,
allowing spectral reconstruction to be performed at lower energies.
There is almost zero bias above 500\,GeV for the \impact{} analysis, with a bias smaller
than 5\% being seen in this stable region for almost all possible observing conditions (0-60$^\circ$ 
 zenith angle and 0-2$^\circ$ source offset). 

 The energy resolution of the \impact{} method
is significantly lower than the \textit{Hillas} methods at all energies, again showing 
the largest improvement at low energies (around 50\% at low energies), where the additional
information used in the fit is most important. Such improved energy resolution should be 
useful when looking for features in a source spectrum, such as the cut-off of a source at high energies,
or the line emission expected from some dark matter annihilation models.

\subsection{X$_{\rm max}$ Resolution}

Figure \ref{fig-Xmax} (left) shows a comparison of the depth of the shower maximum reconstructed
from the likelihood fit with the known X$_{\rm max}$ from the air shower simulation.
The trend in the figure demonstrates that the \impact{} analysis
is able to accurately reconstruct the depth of shower maximum with a small almost constant bias.
In future analyses, the relatively small spread in these reconstructed
values (30 g\,cm$^{-2}$), may help to improve the hadronic 
background rejection at energies below 1TeV.

\subsection{Effective Area}

Effective area 
is defined as the trigger, and subsequent event-selection, probability
multiplied by the area over which simulated gamma-ray showers are scattered.
For current detectors this collection area is much larger 
than the physical footprint of the array, due to the large extent of the 
Cherenkov light pool. This value is generally plotted as a function of energy,
typically rising rapidly at low energies, and being relatively flat high energies,  
sometimes falling slowly at the highest energies as selection cuts
disfavour high-energy events.  

Figure \ref{fig-Xmax} (right) shows the effective area after background rejection cuts
for the \impact{} analysis in comparison with two \textit{Hillas} cut sets.
This figure shows that the effective area of the \impact{} method is very similar to that
of the \textit{Hillas std} method, with a deficit of less than 10\% seen across the full energy range.
The similarity is not surprising as the same
background rejection mechanism is used for both methods
and they are optimised to maximise sensitivity to the same source spectrum.
The effective area of the \textit{Hillas hard} method however has a significantly 
reduced effective area at low energies. This reduction is due to the optimisation of the cuts
using a harder source spectrum, naturally leading to cuts which reject low energy events.

\subsection{Cut Sensitivity}

\begin{figure}[]
\begin{center}

\includegraphics[width=0.49\textwidth]{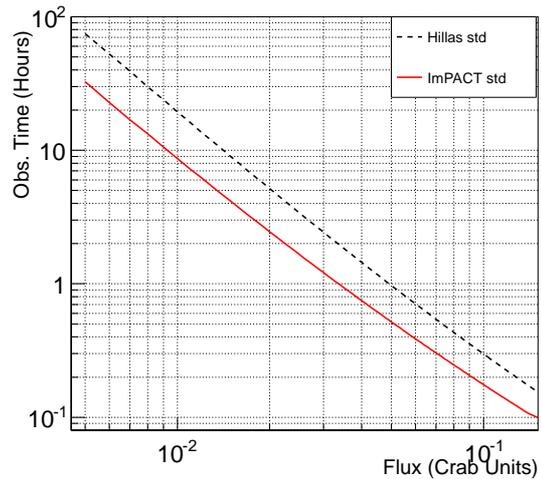}

\caption{Plot of observation time required to detect a point-like
  source of a given flux (expressed as a fraction of the flux of the
  Crab Nebula and assuming a spectrum of the same form: $\frac{dN}{dE} \propto E^{-2.63}$)
  with a statistical significance of 5$\sigma$.}
\label{fig-Sens}
\end{center}
\end{figure}

Figure \ref{fig-Sens} shows the improvements in point source sensitivity resulting
from the increased angular resolving power of the \impact{} analysis. 
Shown is the time 
required to detect a point-like source of gamma-rays with a E$^{-2.63}$ spectrum at 
the 5$\sigma$ level as a function of source strength.  

One can see for the \impact{} method that the time required to detect a source is 
in many cases less than 50\% of that required by the Hillas-based analysis for weak sources. 
For example the observation time required to detect a source of 1\%
Crab strength is around 8.7 hours, in comparison to the 19.3 hours 
required for the Hillas-based analysis with multi-variate background
rejection.

This large improvement in sensitivity is due to the smaller $\theta^2$ cut allowed
by the improved PSF. For the \impact{} analysis standard cuts less than half the 
area of sky used in the Hillas analysis is integrated when observing a point source,
resulting in a corresponding decrease in cosmic ray background events, while the
effective area (and hence number of gamma-rays observed) stays roughly equal.

\section{Source Analyses}

To demonstrate
that the improvement seen in the Monte Carlo simulations 
translates into real world performance, the \impact{} reconstruction was tested
on a strong point-like gamma-ray sources. Comparisons are then made of the source
68\% containment radius against the results of the \textit{Hillas} cuts.

\begin{figure}[]
\begin{center}

\includegraphics[width=0.49\textwidth]{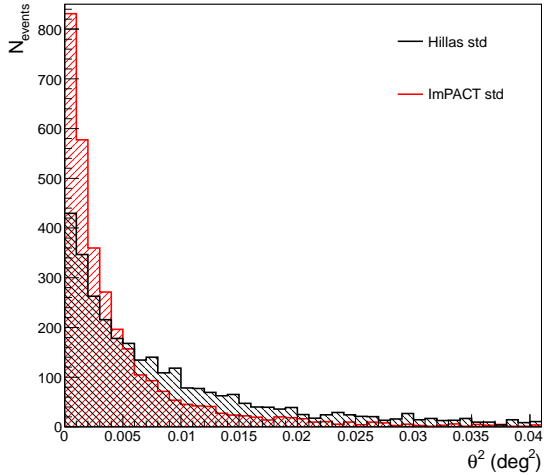}

\caption{Distribution of squared angular distance $(\theta^2$) from
  the source for all events passing background rejection cuts  or a single run 
of PKS\,2155$-$305 flare data using the \textit{Hillas std}  and \impact{}
methods. The background level is approximately
one event per bin in all bins. }
\label{fig-FlareTheta}
\end{center}
\end{figure}

\begin{figure}[]
\begin{center}

\includegraphics[width=0.49\textwidth]{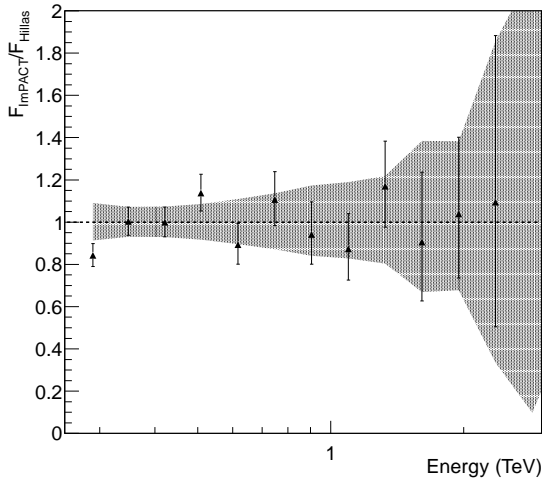}

\caption{Ratio of flux points between ImPACT and Hillas analysis, 
obtained from 14 hours of H.E.S.S. PKS\,2155$-$304 (non-flaring) data. Error bars show calculated 
errors on \impact{} data point, shaded area shows errors on Hillas
points.
}
\label{fig-PKSratio}
\end{center}
\end{figure} 

The ideal data-set for the point spread function comparison is that taken during
the strong flaring activity of the blazar
 PKS\,2155$-$304 \citep{Aharonian2007} in 2006.
At the time, this object was the brightest source of gamma-rays yet
detected by any VHE instrument and the data were taken at small zenith angles, creating a 
large dataset of high-quality gamma-ray events, with over 5900 events post
background rejection cuts remaining in the run used. 

Figure \ref{fig-FlareTheta} show the distribution of excess events in the squared 
angular distance of the reconstructed event position to the test position ($\theta^2$).
This distribution is clearly more peaked toward zero for the \impact{} reconstruction
with a significantly reduced 68\% containment radius.
The improvement in angular resolution seen here, combined with the tighter
angular cut used is able to decrease the number of background events estimated in the 
source region for the \impact{} method, while keeping the estimated signal events
quite similar, increasing the source significance (see table \ref{tab-flare}).

In order to test the energy reconstruction of the \impact{} reconstruction an analysis
was also performed on 14 hours of PKS-2155 observations (while in a non-flaring state)
and an energy spectrum calculated. Figure \ref{fig-PKSratio}
shows ratio of the \impact{} flux points with the points calculated by the Hillas method.
There is very clear agreement between the two sets of flux points across the full energy
range, showing the energy reconstruction performs well on H.E.S.S. data.

\begin{table}[]
\begin{center}
\begin{tabular}{|c|c|c|c|c|}

\hline
Config & N$_{\rm{on}}$& $\alpha$N$_{\rm{off}}$ & $\sigma$ & $\theta_{68}$($^\circ$) \\
\hline
{Hillas std}   & 2290 & 11.9 & 109 & 0.10 \\
{Hillas hard}  & 567   & 1.4 & 58 & 0.074 \\
{ImPACT std} &  2279 & 4.4 & 127 & 0.067 \\
\hline

\end{tabular}
\end{center}
\caption{Event statistics , significance as calculated by the method of \citep{Li1983}  and 68\% event 
containment radius of the first run of PKS\,2155$-$305 flare data, for 3 H.E.S.S cut configurations.}
\label{tab-flare}
\end{table}

\section{Comparison with \textit{model} analysis}

\begin{figure*}[]
\begin{center}
\includegraphics[width=0.49\textwidth]{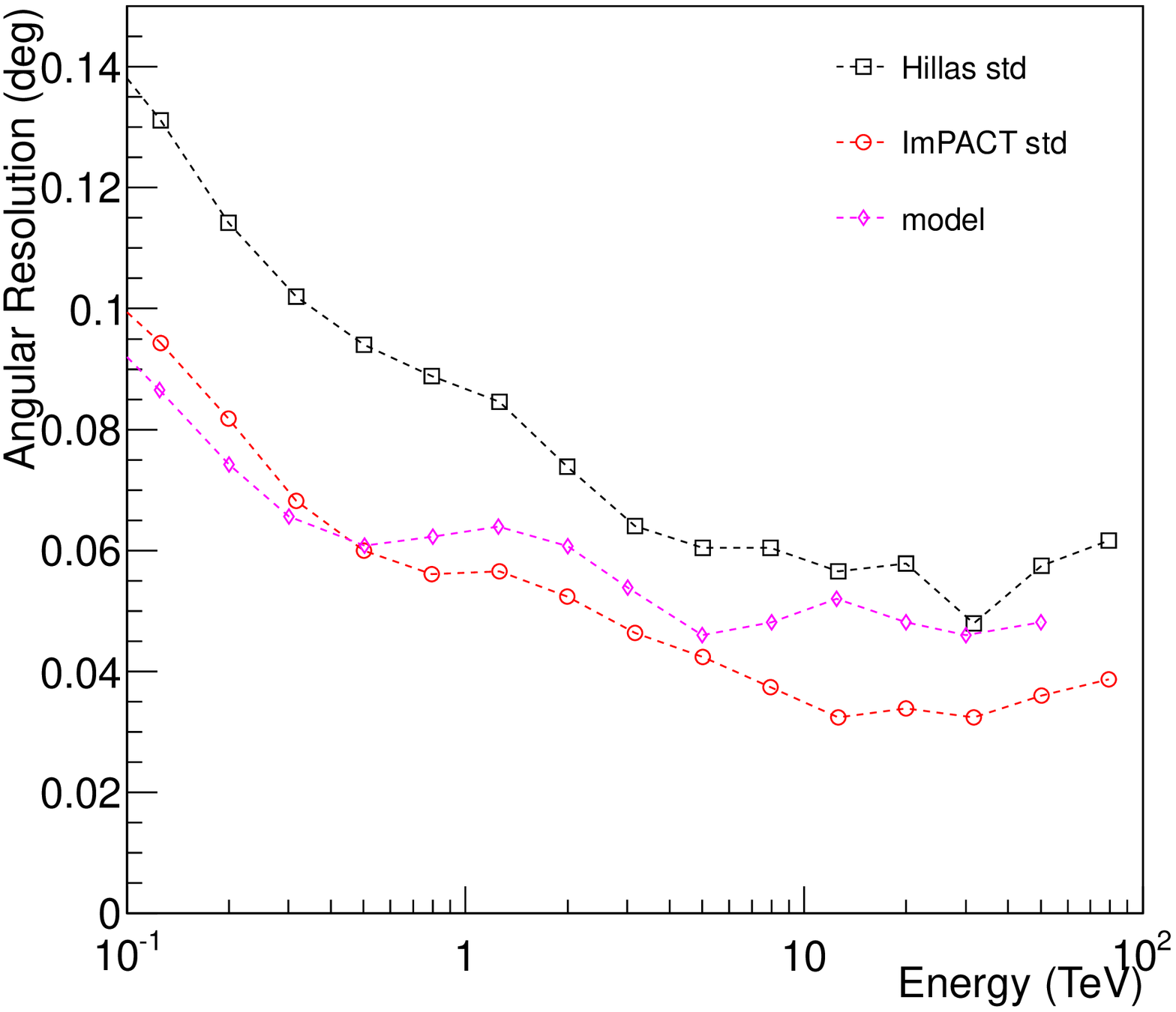}
\includegraphics[width=0.49\textwidth]{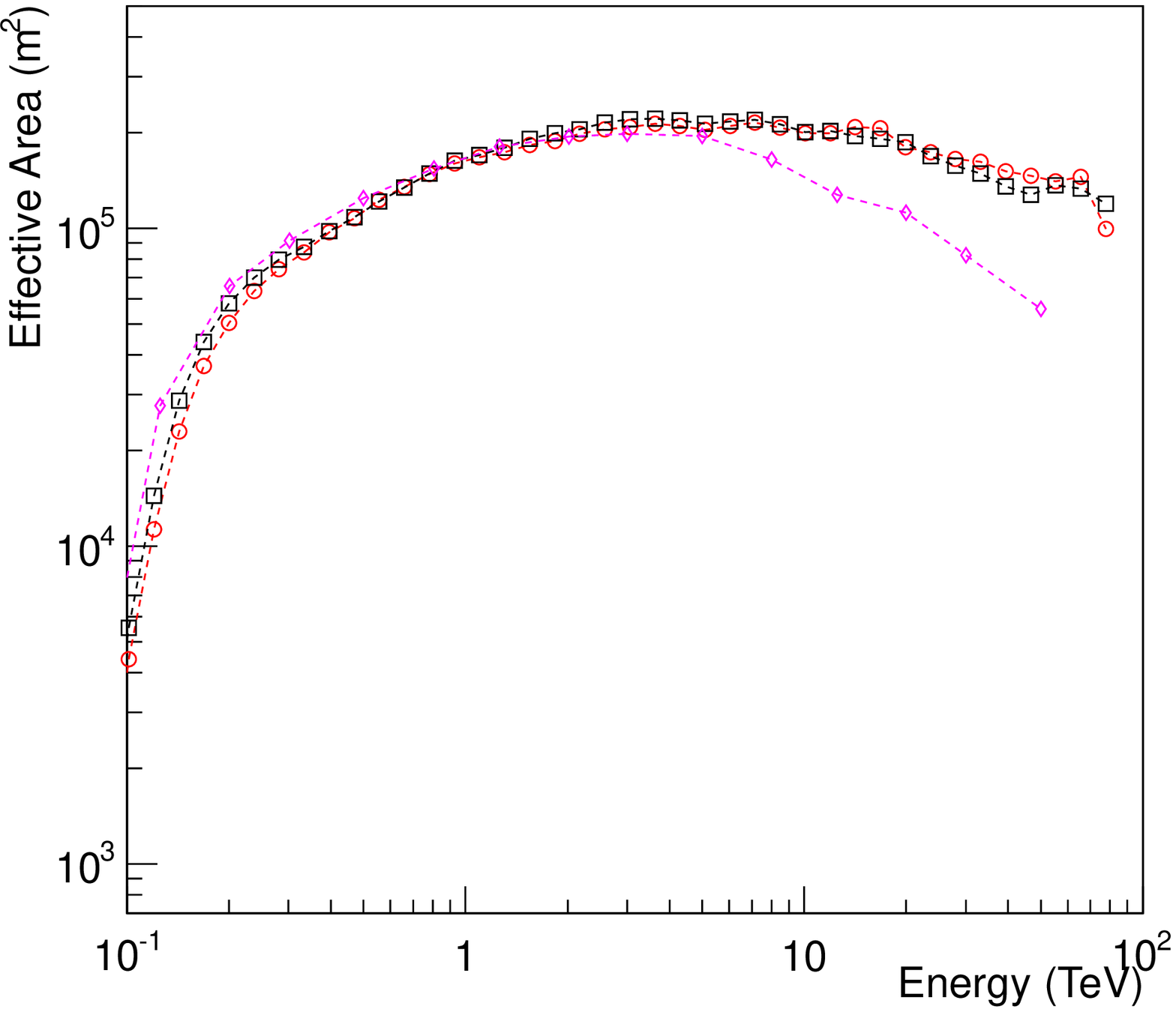}

\caption{Comparison of key performance results from the \impact{} event reconstruction 
with the \textit{model} reconstruction (data points reproduced from
\cite{deNaurois2009}).
Left: Angular resolution (68\% containment radius) as a function of simulated energy. Right: Post cuts effective area
as a function of energy. All curves are calculated at zenith, using standard cuts in all analyses, and HESS simulations with
100\% of the nominal optical efficiency. 
}
\label{fig-ModelComp}
\end{center}
\end{figure*}

The previous section has made detailed comparisons of the \impact{} reconstruction
method with a \textit{Hillas parameter} based method showing large improvements in 
sensitivity, however some comparison to the similar \textit{model} analysis must also 
be made. However, it must be noted that the simulations used in the ImPACT results
and those in the \textit{model} results were produced with different simulation chains
 and there may be some systematic differences between the results of the two chains
 (typically at the 5\% level in total Cherenkov yield \cite{Bernlohr2013}). 

Figure \ref{fig-ModelComp} shows a comparison between the angular resolution
of the \impact{} analysis and the \textit{model} analysis for standard cuts in both analyses. 
It can be 
seen in this plot that at low energies the angular resolution of the \impact{} reconstruction
is similar to that of the \textit{model} analysis. This similarity is expected due to the 
good description of the air shower by the semi-analytical model at low energies and the
relative similarity of the fit procedure. However, at higher energies the angular resolution of the
\impact{} reconstruction is somewhat improved over that of \textit{model}, due to difficultly of
reproducing the behaviour of air showers with the semi-analytical model.
This improvement can also be seen in the post-cuts effective area at high energies, with the
rapid drop-off in effective area demonstrating the difficultly of the \textit{model} analysis to reconstruct
events at high energies, where no such problem exists for \impact{}. Above 10\,TeV
H.E.S.S. observations are typically in the statistics limited regime, such an improvement
in high energy effective area should linearly translate into an improvement in sensitivity.

\section{Conclusions}

We have demonstrated a high performance likelihood-based reconstruction
method for the H.E.S.S. telescope, based on expected image templates
generated from Monte Carlo air shower simulations. This technique 
has been proven to be extremely successful, reconstructing events
with significantly more accuracy than the standard reconstruction algorithm. 
Some improvement has  also been demonstrated over the \textit{model}
reconstruction, especially at the highest energies where semi-analytical modelling 
of the air shower becomes difficult. Use of the \impact{} reconstruction is able to
provide an sensitivity improvement  of around a factor of 2 in observation time
over traditional 
\textit{Hillas-based} reconstruction for point source observations.
Additionally the improvement in angular resolution offers more detailed imaging
of extended sources.

Although some performance improvements have been shown over the 
existing, semi-analytical model based approach to template generation, the 
major advantage of production via MC simulation is the robustness of this approach
against changes in telescope hardware. For example the H.E.S.S. observatory has 
recently commissioned a fifth 26\,m diameter telescope (CT5) at the centre of the array, 
which requires the production of a new set of image templates. 
In order to generate these new templates one can simply re-run the template
generation step using the CT5 simulation configuration, whereas using a model based approach 
the model must be changed to accurately reflect the new hardware. This robustness
will be especially important for the next generation of Cherenkov telescopes, such as the 
Cherenkov Telescope Array, which plan to use multiple telescope types within the same array.

Additionally, this template generation mechanism can in principle be
implemented to allow reconstruction of any particle type. In the case of
protons the large shower to shower fluctuations present may make this 
difficult, however for other particles such as electrons or Iron nuclei
 it may be possible to reconstruct and separate events using this method.

\section*{Acknowledgements}

We thank Prof. C Stegmann, spokesperson for the H.E.S.S. Collaboration and 
Prof G. Fontaine, chairperson of the Collaboration board, for allowing us to use data from the
H.E.S.S. array in this publication. We would like to thank Dr. V. Marandon and Dr. K. Bernl{\"o}hr
for technical assistance in the implementation of the analysis. Finally we would like to thank the 
H.E.S.S. collaboration for useful discussion of the analysis procedure.




\end{document}